**Electrical manipulation of orbital occupancy and magnetic anisotropy in manganites**

*Bin Cui, Cheng Song\*, Gillian A. Gehring, Fan Li, Guangyue Wang, Chao Chen, Jingjing Peng, Haijun Mao, Fei Zeng, and Feng Pan\**

B. Cui, Dr. C. Song, F. Li, G. Y. Wang, Dr. C. Chen, J. J. Peng, H. J. Mao, Dr. F. Zeng, and Prof. F. Pan

Key Laboratory of Advanced Materials (MOE), School of Materials Science and Engineering, Tsinghua University, Beijing 100084, China

Prof. G. A. Gehring

Department of Physics and Astronomy, University of Sheffield, Hicks Building, Sheffield S3 7RH, United Kingdom

E-mail: songcheng@mail.tsinghua.edu.cn; panf@mail.tsinghua.edu.cn

Electrical manipulation of lattice, charge, and spin has been realized respectively by the piezoelectric effect, field-effect transistor, and electric field control of ferromagnetism, bringing about dramatic promotions both in fundamental research and industrial production. However, it is generally accepted that the orbital of materials are impossible to be altered once they have been made. Here we use electric-field to dynamically tune the electronic phase transition in (La,Sr)MnO$_3$ films with different Mn$^{4+}$/(Mn$^{3+}$+Mn$^{4+}$) ratios. The orbital occupancy and corresponding magnetic anisotropy of these thin films are manipulated by gate voltage in a reversible and quantitative manner. Positive gate voltage increases the proportion of occupancy of the orbital and magnetic anisotropy that were initially favored by strain (irrespective of tensile and compressive), while negative gate voltage reduces the concomitant preferential orbital occupancy and magnetic anisotropy. Besides its fundamental significance in orbital physics, our findings might advance the process towards practical oxide-electronics based on orbital.

**1. Introduction**

A wide variety of experimental and theoretical investigations have been pursued intensively to control the performance of materials at the atomic, or even subatomic level for a long period of time.[1–3] The interaction between lattice, charge, spin, and orbital degrees of freedom is found to play a direct and crucial role in the performance of electronic materials, displaying a rich spectrum of exotic phenomena.[4–6] The electric field effect with advantages of low power consumption and high controllability offers an effective and reversible route to confine the lattice, charge, and spin (as illustrated in **Figure 1**): (a) the piezoelectric effect in crystalline materials without inversion symmetry has bridged electric field and lattice for more than 130 years;[7] (b) the field-effect transistor (FET) provided a classic model for the manipulation of carrier density by electrical means, constituting the cornerstone of the semiconductor industry;[8] (c) the magnetization switching is driven reversibly by applied electric fields, which is expected to have a great technological impact on information storage.[9,10] However, electrical control of the missing member—orbital degree of freedom, whose small perturbations would lead to giant responses in the electric and magnetic properties,[11,12] has thus far remained an interesting concept lacking experimental insight.

The orbital represents the shape of electron cloud.[13] In mixed-valence manganites, the occupancy of the $d$ orbital controls the magnitude and anisotropy of the inter-atomic electron-transfer interaction and hence exerts a key influence on the chemical bonding and physical properties.[4,13] For example, orbital occupancy tunes the magnetic behaviors of manganites with the help of strain design, persistently associated with several antiferromagnetic (AF) structures.[14,15] However, the lattice distortion inevitably introduces impurities and extra scattering centers to the system, which are irreversible and hardly controllable, and the orbital occupancy is fixed once the materials have been made and put into use. Application of the field-effect transistor principle to achieve doping variation has modulated the magnetic or electric performances and phase transition in manganites,[16,17] which is expected to provide an approach to tuning orbital occupancy reversibly. The main technological challenge for measuring the change of the orbital order under the influence of an electric field is that the detection of orbital occupancy using x-ray linear dichroism (XLD), calls for an exposed sample with a robust remanent electric field effect,[18] which cannot be achieved by conventionally electrostatic modification of carrier density.[10,11] Fortunately, the relaxation of the gating effect generated by ionic liquid in manganites is negligible even after gate voltage is turned off, guaranteeing *ex situ* measurements of the orbital.[19]

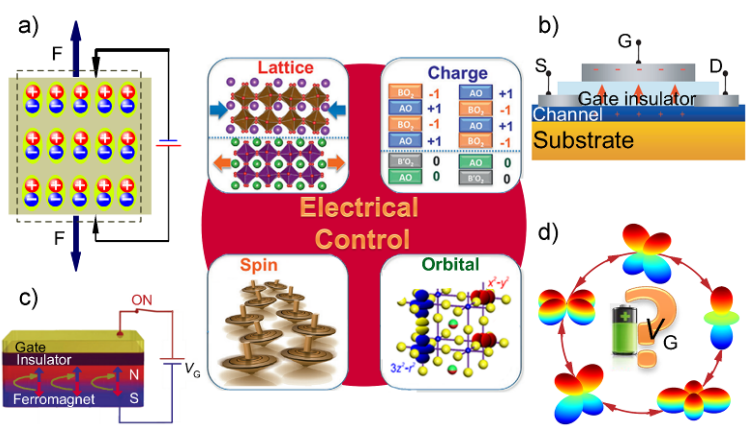

**Figure 1** a) Electric field induced polarization and corresponding lattice deformation in piezoelectric materials; b) The manipulation of carrier density in FET by electric field; c) Electric field controls the magnetization switching in magnetic materials; d) the conception of electrical manipulation of orbital.

## 2. Results and Discussion

A series of 20 nm-thick (La,Sr)MnO$_{3-\delta}$ films with different $x = $ Mn$^{4+}$/(Mn$^{3+}$+Mn$^{4+}$) were prepared by pulsed laser deposition (PLD) at various oxygen background (cooling) pressures of 200 mTorr (300 Torr) ($x$ = 0.54), 100 mTorr (100 Torr) ($x$ = 0.41), and 50 mTorr (10 Torr) ($x$ = 0.20) on STO substrate (Supporting Information Figure S1). The electronic phase diagram of $(La,Sr)MnO_{3-\delta}$ is drawn as a function of $x = Mn^{4+}/(Mn^{3+}+Mn^{4+})$, as shown in **Figure 2**a (Ref. 4), where $x$ could be controlled by the Sr concentration,[4] oxygen pressure during film growth,[18] and gate voltage.[20] We investigate samples of $(La,Sr)MnO_{3-\delta}$ (LSMO) grown with initial values of $x$ = 0.54, 0.41 and 0.20. These are shown as a half-filled triangle, circle and square respectively in Figure 2a. This experimental design locates them close to the boundary between ferromagnetic metallic (FM) and antiferromagnetic metallic or insulating (AFM or AFI) phase, in the centre of the FM phase and at the boundary between FM and insulting [spin-canted insulating (CI) and ferromagnetic insulating (FI)] phase, respectively.

We manipulate the electrical phase transition and Mn $e_g$ orbital occupancy in $(La,Sr)MnO_{3-\delta}$ with different initial $x$ by applying gate voltage ($V_G$) through ionic liquid. Unlike the electrostatic manipulation of carrier density in (Ga,Mn)As (Ref. 21), the electric fields created by the electric double layer (EDL) on a manganite under positive $V_G$ are sufficiently high to change the chemical composition, in this case by driving oxygen ions out of the oxides (Figure 2b).[19,20] This is accompanied by the injection of electrons into films and moves the sample to a lower $x$ as the left arrow shows in Figure 2a. On the contrary, negative $V_G$ extracts electrons with $O^{2-}$ ions migrating back to the films (Figure 2c), increasing the $x$ in LSMO (the right arrow in Figure 2a). When gate voltage is large enough (e.g. +3 V or –3 V), the gating effect based on oxygen migration is stable at room, or lower temperatures with negligible relaxation even after $V_G$ is turned off (Supporting Information Figure S2) which enables us to determine the $x$ values, magnetic and electric properties of samples *ex situ* regardless of the atmosphere. Although the oxygen vacancies migration is not directly observed, the Mn valence variation in XPS and microstructure evolution in HRTEM under different gate voltages suggest the formation and annihilation of oxygen vacancies.[22]

This is also supported by the demonstration of oxygen vacancies-based conduction filaments for resistive switching behaviors, based on the cutting-edge nano-characterization methods[23,24] and theoretical calculations.[25]

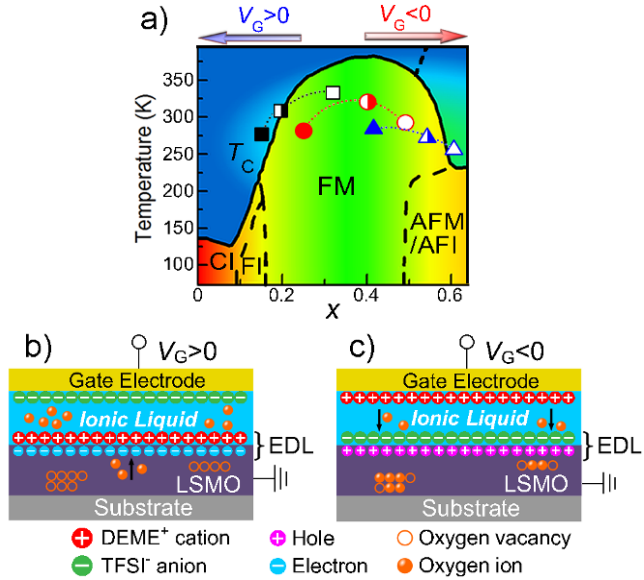

**Figure 2** a) Electronic phase diagram of (La,Sr)MnO$_3$ with varying $x = Mn^{4+}/(Mn^{3+}+Mn^{4+})$ ratio. The half filled triangle, circle, and square denote the positions of samples on STO with initial $x$ = 0.54, 0.41, and 0.20 in phase diagram, respectively. The sample positions after applying positive and negative $V_G$ are marked by corresponding filled and empty symbols. The abbreviations stand for spin-canted insulating (CI), ferromagnetic insulating (FI), ferromagnetic metal (FM), antiferromagnetic metal (AFM), and antiferromagnetic insulating (AFI) phase, respectively. Schematic illustration of the effect of electric double layer (EDL) with b) $V_G > 0$ and c) $V_G < 0$.

We plot the variations of the concentration $x$ determined by XPS (Supporting Information Figure S1), where the full and empty symbols represent the values of $x$ that they reach at voltage of +3 V and −3 V respectively. The heights of the symbols on this plot represent the measured temperature values of the metal-insulator transition as obtained from resistance-temperature (*R-T*) curves (Supporting Information Figure S3). The values of the transition temperatures (equal to Curie temperature, $T_C$) and corresponding $x$ are given in **Table 1** and follow what is expected from the phase diagram. For the case of initial $x = 0.54$, negative $V_G$ drives the sample into the AFM or AFI phase, accompanied by decreased $T_C$, while positive $V_G$ favors the FM phase with the enhancement of $T_C$. When initial $x = 0.41$, the sample is at the FM phase region with optimally magnetic and conductive performances. Hence, both positive and negative voltages result in an abrupt decay of conductivity and $T_C$. In contrast, the magnetic properties of samples with initial $x = 0.20$, are enhanced or suppressed by negative or positive $V_G$, respectively. In conjunction with the growth oxygen pressure and gate voltage, we approximately replicate the dependence of $T_C$ on $x$ in classic phase diagram using a given stoichiometric target. Note that, this phase transition is bulk effect instead of surface carrier accumulation according to the directly microcosmic observation and Mn 2$p$ XPS depth profiling in our previous work.[22] Compared with the electrostatic manipulation based on the capacitance of EDL,[26] the change of $x$ in our system is rather large as the variation of oxygen vacancies is sensitive to the external electric field.

**Table 1** Curie temperature ($T_C$) and the composition ($x$) of tensile strained (La,Sr)MnO$_{3-\delta}$ under various $V_G$.

| $V_G$ (V) | −3 | 0 | +3 |
|---|---|---|---|
| initial $x = 0.54$ (triangles) ($T_C$ / $x$) | 260 K / 0.64 | 271 K / 0.54 | 275 K / 0.43 |
| initial $x = 0.41$ (circles) ($T_C$ / $x$) | 288 K / 0.47 | 315 K / 0.41 | 275 K / 0.25 |
| initial $x = 0.20$ (squares) ($T_C$ / $x$) | 331K / 0.30 | 305 K / 0.20 | 279 K / 0.15 |

We now address the question how the orbital occupancy varies under the electric field. In this part, we concentrate primarily on the sample with $x = 0.54$ because such a nearly half-doped manganite is extremely sensitive to external stimuli, like an electric field.[27] A tensile strain is produced by growing samples on SrTiO$_3$ (STO, $\varepsilon$ = 0.9%) and a compressive strain by growing on LaAlO$_3$ (LAO, $\varepsilon$ = –2.1%). The orbital occupancy is measured by XLD which is the difference between the normalized XAS with photon polarization parallel (E//a) and almost perpendicular (E//c) to the sample plane. The measurement temperature for XLD is 300 K as the interference from ferromagnetism is negligible at a comparatively high temperature, enhancing the accuracy of the XLD signals.[18] The spectra normalization was made by dividing the spectra by a factor such that the $L_3$ pre-edge and $L_2$ post-edge have identical intensities for the two polarizations. After that, the pre-edge spectral region was set to zero and the peak at the $L_3$ edge was set to one. XLD is the difference between the two measurements. Irrespective of the surface symmetry-breaking in such a thick films of 20 nm, the area under XLD around the $L_2$ peak (647.5 eV–660.0 eV, the shadowy areas in Figure 3) could be used to roughly represent the difference between the relative occupancies of $3z^2 - r^2$ [$P(3z^2 - r^2)$] and $x^2 - y^2$ [$P(x^2 - y^2)$] orbital: $A_{XLD} = P(3z^2 - r^2) - P(x^2 - y^2)$ (more details in Supporting Information).[18,28] For tensile strained LSMO without gate voltage, $x^2 - y^2$ orbital is favored with a negative $A_{XLD}$, while the $3z^2 - r^2$ orbital is stabilized ($A_{XLD} > 0$) once compressive strain is applied.

The XAS and the concomitant XLD for the tensile (STO) and compressive (LAO) strained LSMO under representative $V_G$ are presented in **Figure 3**a and b, respectively. The most eminent feature observed here is that the usage of $V_G$ = –3 V, pulls down the absolute value of $A_{XLD}$ while $V_G$ = +3 V does the opposite, as presented in Figure 3 and summarized in **Table 2**. The $x^2 - y^2$ orbital occupancy is strongly weakened by negative $V_G$ but dramatically enhanced by positive one in tensile strained LSMO: $P(x^2 - y^2)$ = 53.1%, 57.8%, and 60.2% for $V_G$ = –3 V, 0 V, and +3 V, respectively. A similar tendency is also found in tensile strained samples of $x$ = 0.41 and 0.20 (Supporting Information Figure S4). When the strain turns to be compressive, $V_G$ of –3 V breaks somehow the $3z^2 - r^2$ orbital occupancy with $P(3z^2 - r^2)$ = 51.3% compared to $P(3z^2 - r^2)$ = 55.1% for $V_G$ = 0 V, while $V_G$ = +3 V stabilizes the $3z^2 - r^2$ orbital with enhanced $P(3z^2 - r^2)$ of 59.3%. It is then concluded that positive or negative voltage strengthens or weakens preferential orbital occupancy, respectively. Although it is difficult to change the orbital occupancy fundamentally and the variations of $A_{XLD}$ are relatively small, the clearly detected signals illustrate the high sensitivity of the orbital occupancy to gate voltage. The magnitude of the variation of orbital occupancy in our case is much larger than that of the calculated value relevant to interfacial ferroelectric polarization.[12] Also, the modulation of $x$ value is uniform from the surface to bottom of film according to the XPS depth profile, which is also supported by the random nucleation model in ionic liquid gated LSMO.[22] Thus the changes in orbital occupancy are almost constant in the whole film rather than limited at the interface or surface.

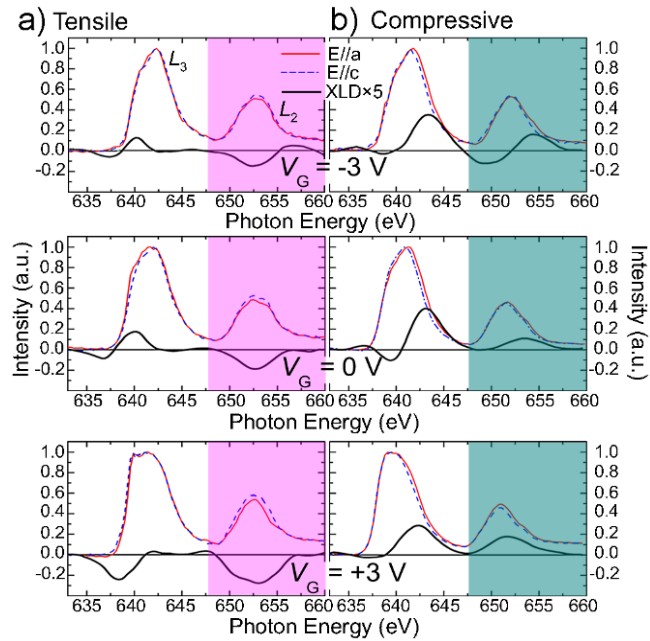

**Figure 3** Normalized XAS [photon polarization parallel (E//a) and almost perpendicular (E//c) to the sample plane] and XLD signals of a) tensile and b) compressive strained LSMO.

The origin of electrical manipulation is the cooperative effect of strain and superexchange. When positive gate voltage is applied, oxygen vacancies in LSMO are increased with injection of electrons and all the samples still locate in the metallic phase, where the $e_g$ doublet is split into a lower and an upper energy orbital due to the strain. The injected electrons fill the orbital starting from the one with lowest energy, that is $x^2 - y^2$ for tensile strain (STO) and $3z^2 - r^2$ for compressive strain (LAO), inducing the enhancement of preferential orbital occupancy. In contrast, negative gate voltage increases the $x$ value with oxygen ions migration into LSMO, which favors the superexchange in doped manganites.[4] The strong superexchange would result in the suppression of electron-transfer interaction and preferential orbital occupancy,[13] corresponding to the small $A_{XLD}$ in Figure 3. Thus, the preferential $e_g$ orbital occupancy is enhanced by positive $V_G$ while broken by negative one, which can be readily generalized to other correlated oxide heterostructures with preferential orbital occupancy, such as $VO_2/TiO_2$ (Ref. 29). We also note that unlike the intrinsic preferential orbital occupancy phase in the classic phase diagram describing the bulk LSMO, the in-plane orbital occupancy of tensile strained samples at $V_G$ = +3 V and –3 V are enhanced and reduced respectively compared to that of the bulk LSMO with the identical $x$, producing a state with a slightly different orbital occupancy from the classic phase diagram.

For $La_{1-y}Sr_yMnO_{3-\delta}$, the Mn can be written as $Mn^{4+}_x$ and $Mn^{3+}_{1-x}$. It is easy to show that $x = y - 2\delta$. In this scenario, introducing oxygen vacancies automatically changes the value of $x$ and hence the phase of LSMO.[30] As a result, the influences of oxygen vacancies $\delta$ and Sr doping concentration $y$ on the performance of LSMO almost overlap, collectively reflecting in the variation of $x$. A closer inspection shows that there are differences in $T_C$ between our samples and the classic phase diagram in Figure 2a: the sample (initial $x$ = 0.54) on STO shows the $T_C$ at $V_G > 0$ and $V_G < 0$ lower and higher than their bulk counterparts, respectively. This difference could arise from two possible origins: additional effect of oxygen vacancies and varied preferential orbital occupancy. Note that positive $V_G$ introduces oxygen vacancies both in the samples with $x$ = 0.20 and 0.54, but results in respectively higher and lower $T_C$ compared with the phase diagram, suggesting that the additional effect of oxygen vacancies should not dominate the performance discrepancy between our samples and the phase diagram, though it cannot be firmly ruled out. Instead, the varied preferential orbital occupancy is most likely the main source responsible for the difference. Additionally, our previous work shows that the $T_C$ of the pristine (La,Sr)MnO$_3$ with in-plane preferential orbital occupancy on STO is around 50 K higher than that on LAO whose preferential orbital occupancy is out of plane, suggesting the importance of orbital occupancy on controlling the magnetic performance.[18]

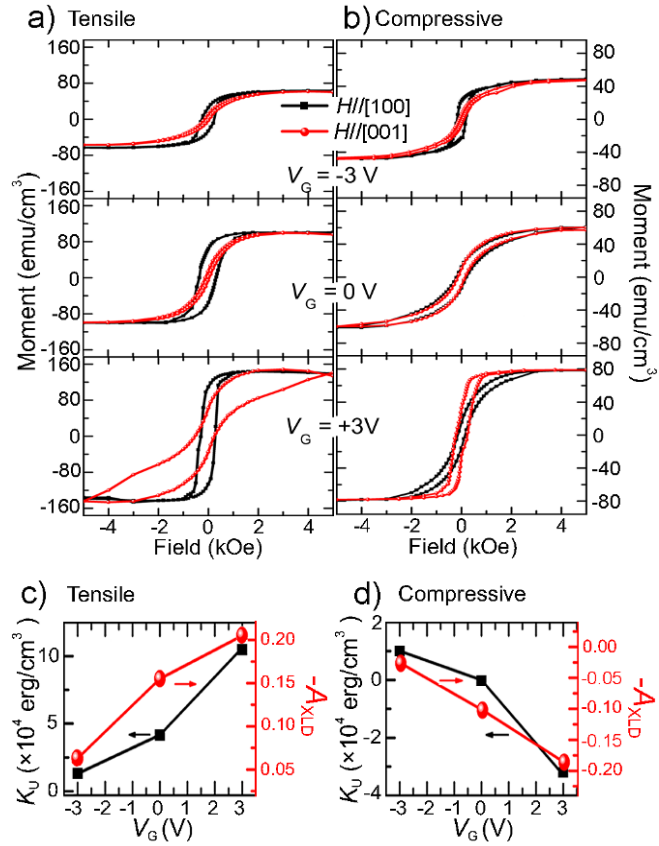

**Figure 4** Magnetization curves measured with field along [100] and [001] directions of LSMO with a) tensile and b) compressive strain under different $V_G$. The dependence of $K_U$ (left axis) and $-A_{XLD}$ (right axis) on $V_G$ for the case of c) tensile and d) compressive strain.

The magnetic switching behavior and magnetic anisotropy under different $V_G$ should be sensitive to the variation in $e_g$ orbital occupancy and carrier density.[15,21,31] We estimate the anisotropy of LSMO by the effective anisotropy constant $K_U = H_A M_S / 2$, where $H_A$ is the difference of the saturated fields between [100] and [001] axes (**Figure 4**a and b), and $M_S$ is the saturated magnetization. The magnetic properties are measured at 10 K, as the form of the magnetic anisotropy is not temperature dependent but its magnitude is increased at low temperatures.[32] Since we do not quantitatively correlate the magnetization and orbital occupancy, the measurements carried out at different temperatures are acceptable. As expected, both shape anisotropy and in-plane orbital occupancy leads the pristine tensile strained sample ($V_G$ = 0 V) to exhibit an in-plane magnetic anisotropy with a $K_U$ of $4.1 \times 10^4$ erg/cm$^3$ (Table 2). By applying gate voltage, only the orbital dependent magnetic anisotropy is manipulated. Positive voltage of +3 V seriously reduces the magnetization switching out-of-plane, resulting in $K_U = 10.6 \times 10^4$ erg/cm$^3$, in contrast to $V_G$ of −3 V corresponding to $K_U = 1.3 \times 10^4$ erg/cm$^3$. The $K_U$ of $V_G$ = +3 V ($x$ = 0.43) is much larger than that of the pristine sample with initial $x$ = 0.41 ($K_U = 5.8 \times 10^4$ erg/cm$^3$), though their compositions are quite close, suggesting that the electrical manipulated $e_g$ orbital occupancy produces a dramatic change in the magnetic anisotropy (Supporting Information Figure S5). When the LSMO changes to be compressive strained (Figure 4b), no detectable magnetic anisotropy is observed in the pristine sample, which can be explained by the competitive result of $e_g$ orbital occupancy and shape anisotropy. This balance is easy to be broken by a perturbation: $V_G$ of −3 V introduces an easy axis in [100] with a $K_U$ of $1.0 \times 10^4$ erg/cm$^3$, while $V_G$ = +3 V makes the film behave as perpendicular magnetic anisotropy with a negative $K_U$ of $-3.3 \times 10^4$ erg/cm$^3$. Remarkably, combinations of $V_G$ dependent $-A_{XLD}$ and $K_U$ in Figure 4c and d confirm that the increased (decreased) in-plane orbital occupancy at positive (negative) $V_G$ strengthens (weakens) the magnetic anisotropy of LSMO with the [100] easy-axis, while the enhanced $3z^2 - r^2$ orbital occupancy in compressive strained LSMO as $V_G$ changes from −3 V to +3 V would change the easy axis of LSMO from in-plane to out-of-plane gradually.

In addition to magnetic anisotropy measurements we have explored gate voltage dependent saturation magnetization, which is not so robust because the films under various $V_G$ are close the boundary between FM and AFM/AFI phases. Positive $V_G$ of +3 V drives the pristine sample on STO with $x > 0.4$ from the AFM or AFI scale to the FM phase region in phase diagram (Figure 2a), as the injection of electrons favors the double exchange effect, apparently enhancing the $M_S$ though the preferential $e_g$ orbital occupancy is increased in this case. Corresponding values for $x = 0.54$ are listed in Table 2. In stark contrast, for the case of –3 V, extraction of electrons pulls the sample deeper into the antiferromagnetic phase with ferromagnetism suppression. For compressive strained LSMO, the magnetization exhibits a similar dependence on $V_G$ that $M_S$ = 47.2 emu/cm$^3$, 59.2 emu/cm$^3$, and 78.4 emu/cm$^3$ for $V_G$ = –3 V, 0 V, and +3 V, respectively. It is generally accepted that the $e_g$ orbital occupancy induced by the strain effect in manganite films inevitably deteriorates the ferromagnetic ordering and conductivity, which has been an obstacle for the integration of manganites into spintronics.[15] The increased preferential $e_g$ orbital occupancy associated with magnetization and conductivity enhancements for electrical manipulation of ferromagnetism in the present case reaffirms the difference between electrically manipulated and bulk LSMO with the same $x$, providing a promising way for the introduction of strong magnetic anisotropy without the cost of relevant magnetic and electric properties. It is useful to point out that the enhancement of magnetic anisotropy would reduce the width of domain wall, likely enhancing the domain wall resistance (more details in Supporting Information).[33–35]

The relationship between the variations of orbital occupancy, magnetic anisotropy, and magnetization could be understood from the perspective of orbital mixing. The measurements of the XLD and the values of orbital occupancy were made at room temperature when the samples were above, or very close to, their ordering temperature, where the carriers in the samples will be polarons. Thus the $e_g$ band states would be combinations of the states given by Equations 1 and 2 (Ref. 36):

$$|\psi_{low}\rangle = \cos\frac{\theta}{2}|3z^2 - r^2\rangle + \sin\frac{\theta}{2}|x^2 - y^2\rangle \qquad (1)$$

$$|\psi_{high}\rangle = \sin\frac{\theta}{2}|3z^2 - r^2\rangle - \cos\frac{\theta}{2}|x^2 - y^2\rangle \qquad (2)$$

where $|\psi_{low}\rangle$, $|\psi_{high}\rangle$, $|3z^2 - r^2\rangle$, and $|x^2 - y^2\rangle$ are the wave functions of low, high energy states, $3z^2 - r^2$, and $x^2 - y^2$ orbital, respectively. The mixed orbital is the hybridization of $3z^2 - r^2$ and $x^2 - y^2$ orbital, whose proportions $P(3z^2 - r^2)$ and $P(x^2 - y^2)$, respectively, are obtained from XLD results. Then the angle of orbital mixing for low energy state can be calculated from $P(3z^2 - r^2) = \cos^2\frac{\theta}{2}$ and $P(x^2 - y^2) = \sin^2\frac{\theta}{2}$, according to the normalization and orthogonality of orbital hybridization. In manganites, the magnetization is determined by the band width while the electron transfer integrals depend on the mixing angle $\theta$ (Ref. 37). For an unstrained magnetite in the metallic regime $\theta = 90°$, so that $P(x^2 - y^2) = P(3z^2 - r^2)$; this corresponds to maximum transfer integrals between the Mn ions. In the films grown on STO which are under tensile strain the nearest neighbors separation will be greater in the plane.

The values of $\theta$ measured in our experiments, shown in Table 2, are all relatively close to 90°. The $\theta$ values of films grown on STO are all larger than 90°, corresponding to a higher probability $P(x^2 - y^2)$, while those grown on LAO have a higher probability $P(3z^2 - r^2)$ with $\theta < 90°$. Thus, for the orbital where the nearest neighbor separations are reduced by the strain, their occupations are enhanced, in-plane for tensile strain (STO) and out-of-plane for compressive strain (LAO). Positive $V_G$ introduces electrons into the orbital enhanced by strain, giving rise to a larger value of $|\theta - 90°|$. While negative $V_G$ increase the $x$ value with stronger superexchange effect, which drives the value of $\theta$ close to 90°. It is also clear from this table that the magnitude of $|\theta - 90°|$ increases as the saturation magnetization increases. In the ferromagnetic phase, much of the stabilization energy of the magnetization comes from the increase in the band width in the partially occupied $e_g$ band due to efficient electron transfer between Mn ions facilitated by the parallel spin electrons in the Mn $t_{2g}$ shell.[4] This increase in $|\theta - 90°|$ corresponds to an enhancement of the transfer integrals along the directions where the distortions are reducing the Mn–Mn spacing. Hence we can understand the orbital enhancement in these films in terms of the benefits of increasing the transfer integrals when the spin ordering allows this effect to increase the band width. The induced preferential orbital occupancy naturally impacts on the magnetic anisotropy. The enhancement of $P(x^2 - y^2)$ as the voltage reduces the value of $x$ from 0.64 to 0.43 leads to a strong uniaxial anisotropy for the film grown on STO under a voltage of +3 V. A different effect is seen for the film grown on LAO: the anisotropy is in-plane for a voltage of –3 V but this is reduced to zero at 0 V and then becomes out-of-plane as +3 V is applied with increased magnetization.

**Table 2** The composition ($x$), $A_{XLD}$, magnetic properties, and orbital mixing angle $\theta$ of tensile and compressive strained LSMO (initial $x$ = 0.54) at various $V_G$.

| | Tensile | | | Compressive | | |
|---|---|---|---|---|---|---|
| $V_G$ (V) | –3 | 0 | +3 | –3 | 0 | +3 |
| $x$ | 0.64 | 0.54 | 0.43 | 0.63 | 0.53 | 0.43 |
| $A_{XLD}$ | –0.063 | –0.155 | –0.205 | 0.026 | 0.102 | 0.186 |
| $M_S$ (emu/cm$^3$) | 61.6 | 99.2 | 143.2 | 47.2 | 59.2 | 78.4 |
| $K_U$ (×10$^4$ erg/cm$^3$) | 1.3 | 4.1 | 10.6 | 1.0 | 0.0 | –3.3 |
| $\theta$ (°) | 93.66 | 98.97 | 101.88 | 88.52 | 84.17 | 79.29 |

## 3. Conclusions

In conclusion, an electrical manipulation of phase transition and orbital occupancy in LSMO is realized by applying gate voltage in a reversible and quantitative manner, generating a discrepant state with different orbital occupancy but the same Mn$^{4+}$/(Mn$^{3+}$+Mn$^{4+}$) compared to the bulk. For negative $V_G$, the extraction of electrons drives LSMO into the AFM/AFI phase accompanied with uniform orbital occupancy, orbital mixing angle close to 90°, and reduced magnetic anisotropy. On the contrary, when $V_G$ is positive, electrons are injected into strain-stabilized orbital of LSMO, driving the mixing angle away from 90°. For the manganite films near the antiferromagnetic phase, positive gate voltage could simultaneously enhance the magnetic anisotropy and magnetization. In the case of preferential orbital occupancy and the concomitant magnetic anisotropy we have seen that the control of $x$ using voltage has additional benefits (simultaneous higher magnetization and anisotropy) more than could be obtained by growth parameters alone. Although the difficulty of handling ionic liquid, our findings present a broad opportunity to advance the control of orbital in a variety of correlated electron system.

## 4. Experimental Section

*Sample preparation:* 20 nm-thick $(La,Sr)MnO_{3-\delta}$ films were grown using PLD from a stoichiometric $La_{0.6}Sr_{0.4}MnO_3$ target by applying a KrF excimer laser at a rate of 0.77 nm/min. STO ($a_{STO}$ = 3.90 Å) and LAO ($a_{LAO}$ = 3.79 Å) substrates are used to introduce tensile and compressive strain, respectively. The growth was monitored in situ by RHEED (reflection high-energy electron diffraction) analysis allowing precise control of the thickness at the unit cell scale and accurate characterization of the growth dynamics. The $Mn^{4+}/(Mn^{3+}+Mn^{4+})$ ratios of films are controlled by the oxygen pressure during growth and cooling process: when the growth (cooling) oxygen pressures are 200 mTorr (300 Torr), 100 mTorr (100 Torr) and 50 mTorr (10 Torr), the $Mn^{4+}/(Mn^{3+}+Mn^{4+})$ ratios are 0.54, 0.41, and 0.20, respectively.

*Sample characterization:* Ionic liquid N,N-diethyl-N-(2-methoxyethyl)-N-methylammonium bis-(trifluoromethylsulfonyl)-mide (DEME-TFSI) with the freezing temperature of around 210 K was used as the electrolyte, whose irreversible reaction with manganites has been carefully ruled out (Ref. 38, Supporting Information Figure S6 and S7). Samples with large areas of 2.5 × 2.5 mm$^2$ were entirely covered by ionic liquid with Au top electrodes and a specific voltage is applied between the Au electrode and (La,Sr)MnO$_3$ film by Agilent 2901A for about 30 min without special instruction. The values of *x* are determined by x-ray photoelectron spectroscopy (XPS) at 300 K. XLD characterizations were performed in total electron yield (TEY) mode at Beamline BL08U1A at Shanghai Synchrotron Radiation Facility. A superconducting quantum interference device (SQUID) was used to measure magnetic properties at 10 K. Before XPS, XLD, and magnetization measurements, the ionic liquid is removed by rinsing in acetone and alcohol. Conductivity was probed in a transistor structure (Supporting Information Figure S3) by the physical property measurement system (PPMS).

**Supporting Information**

Supporting Information is available from the Wiley Online Library or from the author.

**Acknowledgements**

The authors are grateful to Prof. P. Yu for fruitful discussions and critical reading of manuscript. We acknowledge Beamline BL08U1A in Shanghai Synchrotron Radiation Facility (SSRF) for XAS/XLD measurements and Center for Testing & Analyzing of Materials for technique support. This work was supported by the National Natural Science Foundation of China (Grant Nos. 51322101, 51202125 and 51231004) and National Hi-tech (R&D) project of China (Grant no. 2014AA032904 and 2014AA032901).